\documentclass[10pt,aps,prl,onecolumn,superscriptaddress,floatfix,nofootinbib,notitlepage]{revtex4-1}
\usepackage{amsmath,amssymb,amsfonts}
\usepackage{graphicx,color}
\usepackage{natbib}
\usepackage{hyperref}
\newcommand{\ie}{\emph{i.e.}}
\newcommand{\eg}{\emph{e.g.}}

\newcommand{\avg}[1]{\langle #1\rangle}

\begin{document}

\title{Unfolding the innovation system for the development of countries:\\co-evolution of Science, Technology and Production}

\author{Emanuele Pugliese}\email{emanuele.pugliese@gmail.com}
\affiliation{Istituto dei Sistemi Complessi (ISC)-CNR, 00185 Rome - Italy}
\affiliation{International Finance Corporation, World Bank Group, 20433 Washington - USA}
\thanks{The findings, interpretations, and conclusions expressed in this paper are entirely those of the authors. 
They do not necessarily represent the views of the International Bank for Reconstruction and Development/World Bank and its affiliated organizations, 
or those of the Executive Directors of the World Bank or the governments they represent.}
\affiliation{University of Bath, Bath BA27AY - United Kingdom}
\author{Giulio Cimini}
\affiliation{IMT School for Advanced Studies, 55100 Lucca - Italy}
\affiliation{Istituto dei Sistemi Complessi (ISC)-CNR, 00185 Rome - Italy}
\author{Aurelio Patelli}
\affiliation{Istituto dei Sistemi Complessi (ISC)-CNR, 00185 Rome - Italy}
\author{Andrea Zaccaria}
\affiliation{Istituto dei Sistemi Complessi (ISC)-CNR, 00185 Rome - Italy}
\affiliation{International Finance Corporation, World Bank Group, 20433 Washington - USA}
\thanks{The findings, interpretations, and conclusions expressed in this paper are entirely those of the authors. 
They do not necessarily represent the views of the International Bank for Reconstruction and Development/World Bank and its affiliated organizations, 
or those of the Executive Directors of the World Bank or the governments they represent.}
\author{Luciano Pietronero}
\affiliation{Dipartimento di Fisica, Sapienza Universit\`a di Roma, 00185 Rome - Italy}
\affiliation{International Finance Corporation, World Bank Group, 20433 Washington - USA}
\thanks{The findings, interpretations, and conclusions expressed in this paper are entirely those of the authors. 
They do not necessarily represent the views of the International Bank for Reconstruction and Development/World Bank and its affiliated organizations, 
or those of the Executive Directors of the World Bank or the governments they represent.}
\affiliation{Istituto dei Sistemi Complessi (ISC)-CNR, 00185 Rome - Italy}
\author{Andrea Gabrielli}
\affiliation{Istituto dei Sistemi Complessi (ISC)-CNR, 00185 Rome - Italy}

\begin{abstract}
We show that the space in which scientific, technological and economic developments interplay with each other 
can be mathematically shaped using pioneering multilayer network and complexity techniques. 
We build the tri-layered network of human activities (scientific production, patenting, and industrial production) and study the 
interactions among them, also taking into account the possible time delays.
Within this construction we can identify which capabilities and prerequisites are needed to be competitive in a given activity, 
and even measure how much time is needed to transform, for instance, the technological know-how into economic wealth and scientific 
innovation, being able to make predictions with a very long time horizon. Quite unexpectedly, we find empirical evidence that the naive knowledge 
flow from science, to patents, to products is not supported by data, being instead technology the best predictor for industrial and scientific
production for the next decades.
\end{abstract}


\maketitle

Knowledge production and organization represents the main activity of modern societies -- 
``learning economies'' \cite{Lundvall1994} in which most of the wealth of a country 
is intangible, and the organization of the national innovation system \cite{Nelson1993}, 
and of diffused creativity \cite{Florida2014} are the crucial capabilities for success. 
Therefore, in the last thirty years the relationships between science, technology and economic competitiveness 
has become an important focus for social sciences in general and economics in particular \cite{Romer1990,Dosi1988}.
Even though the standard narrative links science, technology and economic productivity 
in a direct flow \cite{Bozeman2000}, actual interactions are likely multi-directional \cite{Hughes1986} and 
emerging from the non-trivial interplay among their individual components: the footprint of a complex system.

The new literature of Economic Complexity uses techniques which,
differently from traditional social science approach, do not try to average out the complexity of the system, 
but embraces it by explicitly building on the heterogeneity of individual actors, activities and interactions to
extract relevant parameters to characterize the system.
Trying to recover the qualitative insights \cite{Hirschman1988} and the few quantitative 
attempts \cite{Teece1994,Nesta2005} of the heterodox economists and social scientists,
researchers used this approach to study unobservable characteristics and capabilities of countries 
\cite{Hidalgo2009,Tacchella2012,Cimini2014}, and to unearth unexpected synergies 
among different activities \cite{Hidalgo2007,Zaccaria2014}.

Following this line \cite{Teece1994,Hidalgo2007,Zaccaria2014,klimek2012empirical}, 
here we create the network of interactions between the different human activities. 
We build on the assumption that if two activities co-occur significantly more often than randomly (in terms of appropriate null models) in the same countries at the same time, 
then there is an overlap between the capabilities required to achieve proficient level (\ie, competitive advantage) in both.
For the first time our network encompasses activities in different realms (or \emph{layers}): scientific fields, technological sectors, and economic production. 
In such a comprehensive \emph{multi-layer} network \cite{Boccaletti2014}, 
\ie, a system where entities belong to different sets and several categories of connections exist among them, 
we particularly focus on interactions \emph{among} the different layers.
As detailed in the Supplementary Information, to task we build the adjacency matrices 
$M^L_{c,a}(y)$ connecting country $c$ with the activity $a$ belonging to layer $L$ if, in year $y$, 
country $c$ was expressing a competitive advantage in activity $a$ with respect to the world average. 
$L$ stands for the layer of analysis, consisting in the set of all activities related to either $S$cience, $T$echnology or $P$roducts export. 
Note that each of these layers has an intrinsic hierarchical structure: for instance, in the science layer we can consider activities like \emph{Physics and Astronomy} 
or corresponding sub-activities (like \emph{Statistical Physics, Condensed Matter Physics, Nuclear and High Energy Physics}). 
Thus, our matrices do depend on the resolution used for activities classification (even if not explicitly reported in the notation). 
We use different established databases to construct the multi-layer space: 
for \emph{S}cience, we take bibliometric data on papers in the period in the various scientific fields from Scopus (\url{www.scopus.com});
for \emph{T}echnology, we consider the number of patents in different technological sectors extracted from Patstat (\url{www.epo.org/searching-for-patents/business/patstat}); 
and for \emph{P}roducts export, we use the export data collected by UN COMTRADE (\url{https://comtrade.un.org/}).

\begin{figure}
\includegraphics[width=0.85\textwidth]{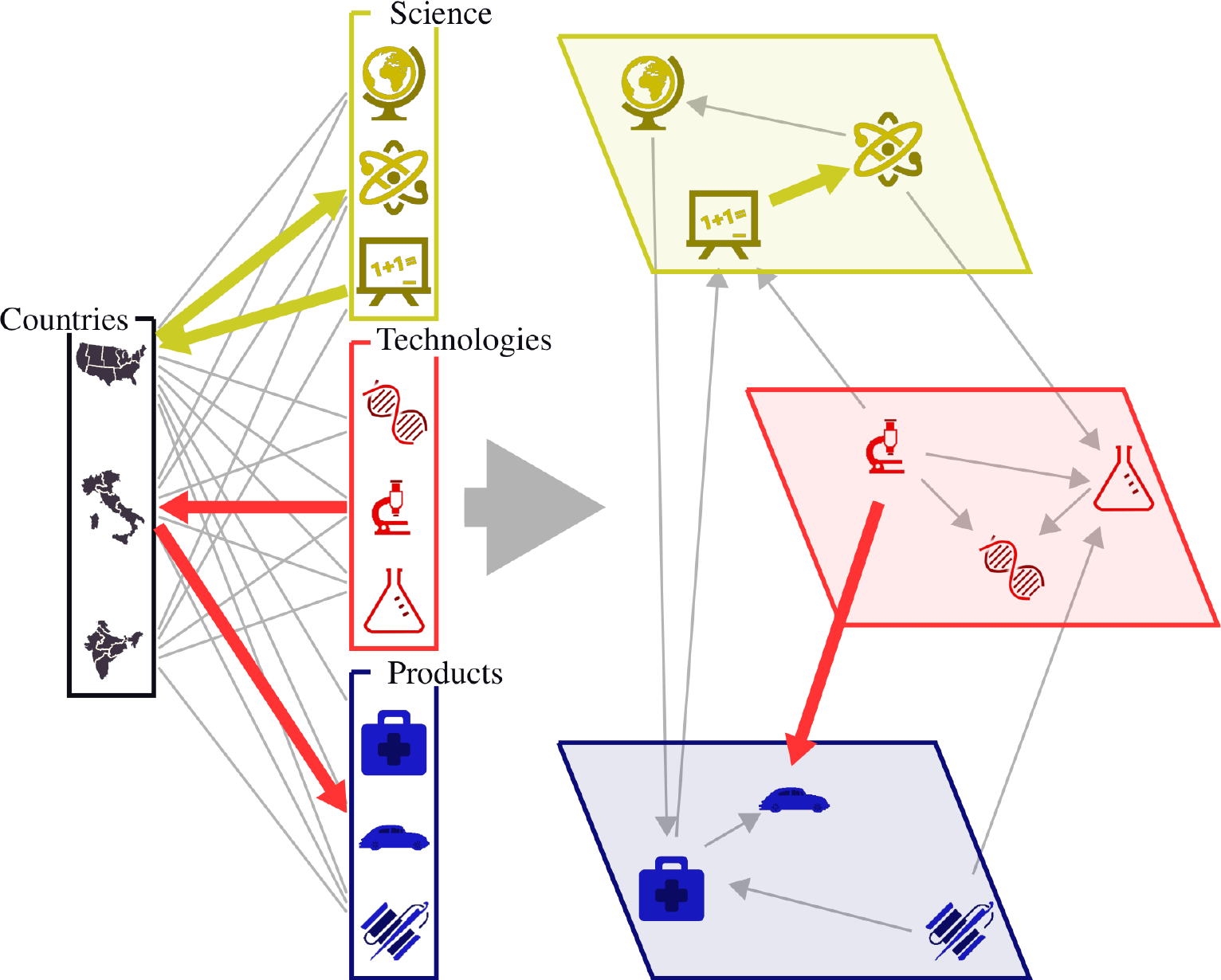}
 \caption{{\em Left panel}: schematic visualization of the triple bipartite network with \emph{C}ountries in one partition and activities (\emph{S}ciences, \emph{T}echnologies, and
 \emph{P}roducts) in the other one; {\em Right panel}: tri-layer representation of the resulting \emph{Assist} matrix between activities. 
 The generic element of the \emph{Assist} matrix is equal to the probability that  a bit of information, randomly diffusing in the triple bipartite network, travels from one activity to another. 
 This can happen in the same activity layer, as it is the case for the yellow path linking two sciences, or intra-layer, as  it is the case for the red path going from a technology to a product.}
 \label{fig:probability}
\end{figure}

Using these matrices we compute the probability of having a comparative advantage in activity $a_2\in L_2$ in the year $y_2$, 
conditional to having a comparative advantage in activity $a_1\in L_1$ in the year $y_1$ (Figure \ref{fig:probability}), that we define as the \emph{Assist} matrix, $B$: 
\begin{align*}  
B_{a_1\to a_2}^{L_1\to L_2}(y_1,y_2)=&Pr(a_2;y_2|a_1;y_1)=\sum_c Pr(a_2;y_2|c)Pr(c|a_1;y_1)=\\
  =&\sum_c{\frac{M_{c,a_2}^{L_2}(y_2)}{d^{L_2}_{c}(y_2)}}\frac{M_{c,a_1}^{L_1}(y_1)}{u^{L_1}_{a_1}(y_1)} ,\\
\end{align*}
where $u_a=\sum_c M_{c,a}^L$ is the ubiquity of activity $a\in L$, and $d_c^L=\sum_{a\in L} M_{c,a}^L$ is the diversification of country $c$ in the layer $L$.
Note that we assumed that countries discount all information 
about capabilities, \ie, that $Pr(a_2;y_2|c,a_1;y_1)=Pr(a_2;y_2|c)$.
These probabilities can be associated with the overlap between the capabilities required to perform activities $a_1$ in the year $y_1$ and $a_2$ in the year $y_2$, 
under the assumption that the capability structure is not country specific. 
In order to interpret the numbers and assess their significance, we have to compare them with a null model, 
which we obtain adopting a novel algorithm to randomize bipartite networks \cite{Saracco2015,Gualdi2016,Saracco2016}. Thus, for each pair of (ordered) activities and years, 
we have an empirical observed value of $B_{a_1\to a_2}^{L_1,L_2}(y_1,y_2)$ 
and a distribution of probability for such value under the null hypothesis 
that the ubiquity of activities and diversification of countries in each layer sum up all the information. 
This means that activities are independent, and there is no capability structure behind the networks. 
Therefore co-occurrences between activities happens at random, some more likely than others only because of the 
ubiquities of the two activities -- \eg, some technological fields and the export of some products
are less ubiquitous and therefore they are both more likely to be performed by advanced and more diversified countries.
Any specific link $a_1\to a_2$ for which we can reject such null hypothesis is interpreted as the signal of some real 
interdependency between the specific capabilities required by a country to perform those specific activities -- at the same time or with a time delay.

\begin{figure}
\includegraphics[width=0.9\textwidth]{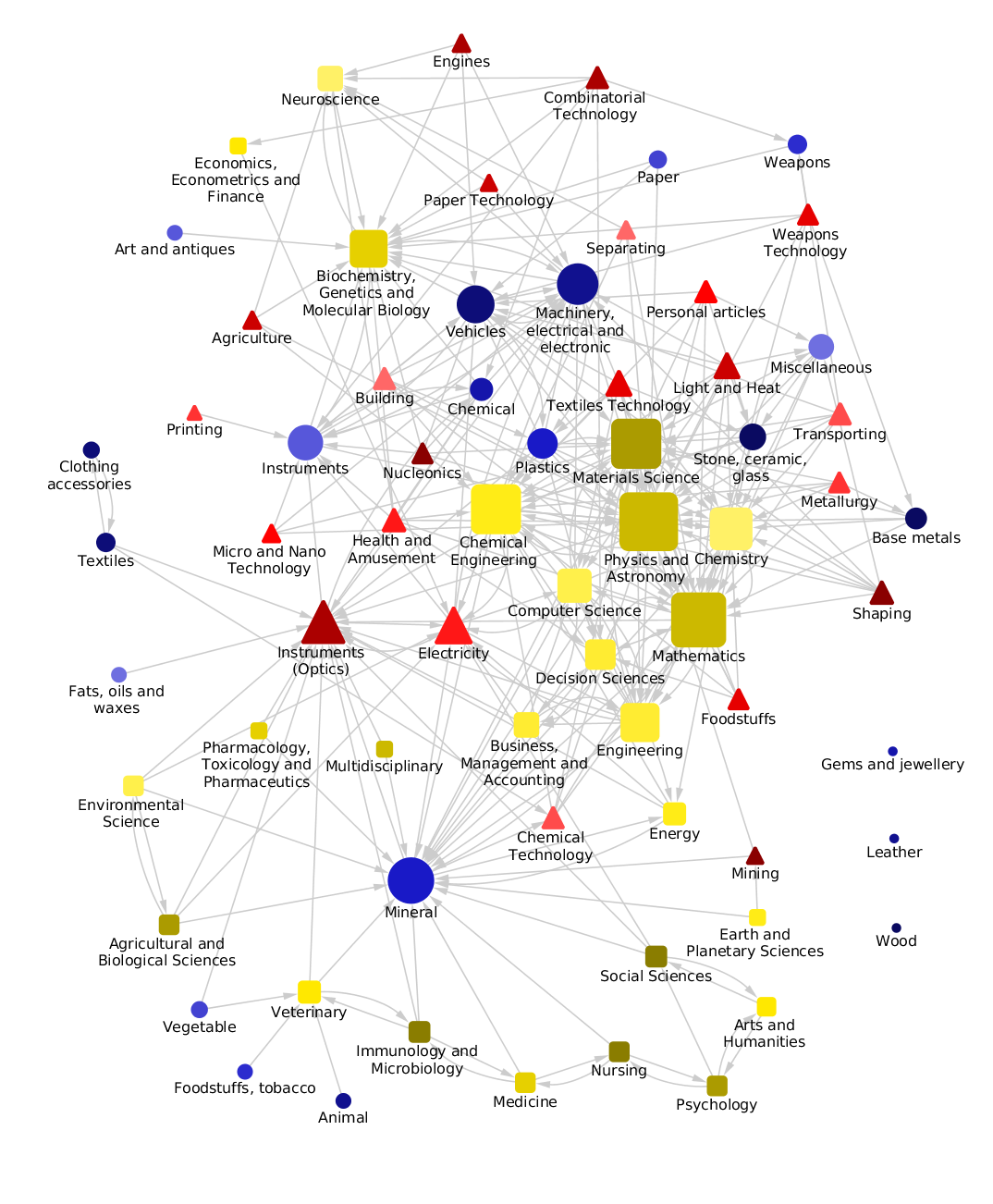}
 \caption{The multilayer network of broadly aggregated activities (23 scientific major categories, 25 technological sub-sections, 21 product sections). The links are
 obtained using a significance level of 99.999\%. To increase the signal-to-noise ratio, we compute $B$ as the average of three consecutive years in the middle of our sample (2008-2010). 
 Red triangular nodes represent technologies, yellow squared nodes represent scientific fields and finally blue circle nodes represent the export of products. The node sizes are proportional to the degree.}\label{fig:Network}
\end{figure}

\begin{figure}
\begin{center}
\includegraphics[width=\textwidth]{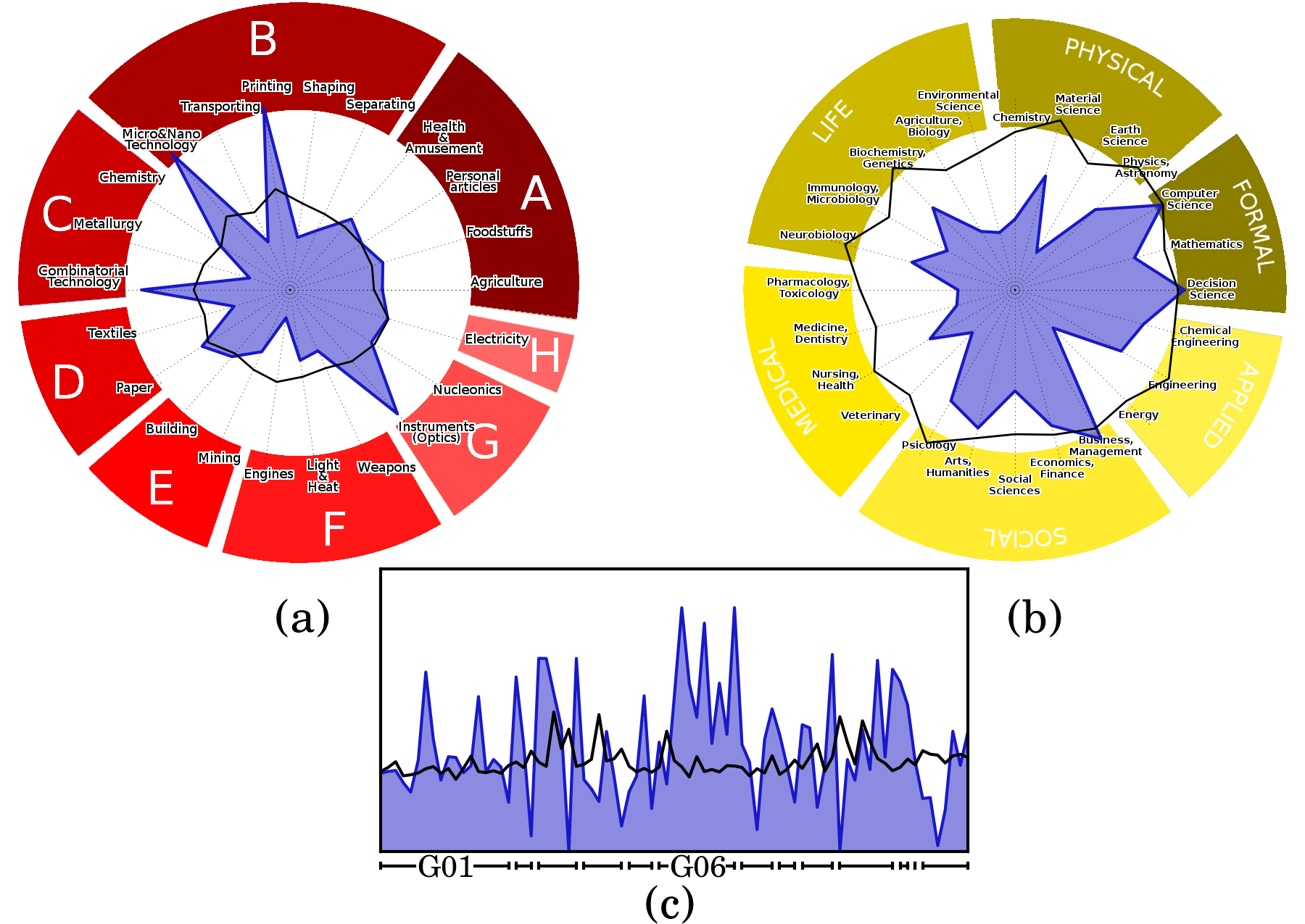}
\caption{Technology fields (a) and science fields (b) in 2004-2008 correlated with 
successful export of Desktop Computers (Harmonized System code 847149) in 2006-2010. 
The blue contour corresponds to the empirical values of $B$, the black line denotes the 95\% confidence interval. 
(c) Higher resolution analysis with technological section ``G: Physics'' expanded in its sub-classes, the black line denoting again the 95\% confidence interval.
The peak in ``G06'' corresponds to the ``Computing'' class.
\label{fig:oneProduct}}
\end{center}
\end{figure}

The described methodology allows obtaining unprecedented qualitative and quantitative insights on the complex dynamics of development. 
By linking together those activities which are related at a given significance level, 
we can build the whole multilayer space in  which scientific, technological and export activities are embedded (Figure \ref{fig:Network}). 
We also perform a more focused analysis and show how a detailed activity (\eg, the export of an individual product) 
is related to activities in other layers at various aggregation levels. An example is shown in Figure \ref{fig:oneProduct}, 
where we plot the scientific and technological fields related to the export of Desktop computers.
We can draw from the figure two observations: i) significant peaks, \ie, events observed in the real data with less than 5\% probability to occur in the null model, 
are meaningful according to our understanding of the scientific and technological prerequisites to be competitive in Desktop Computer export; 
ii) technology tend to be more significantly related to export than science, the signal-to-noise ratio being higher. 
This is not an exception related to this product, as we shall see next.

The representation of Figure \ref{fig:Network} suggests that any modeling of the Innovation System 
assuming a standard direction of the dynamics between layers (e.g., from Science to Technology to Products) is
simplistic. In other words the assumption that all activities belonging to the same layer behaves similarly in determining the innovation cascade is empirically unjustified: 
the division in Scientific, Technological and Production activities is not greatly informative of their role in the network. 
A Technological activity could be a precursor of some Scientific activities, 
while the opposite could be true for different activities in the same layers: a complex system where no set of labels is 
fully informative of the dynamic structure.
We can however provide some insights on the average interactions between activities in two specific layers to assess if the aggregate models of the innovation 
systems\cite{Lundvall1994} offer at least a proper description of the most common modes of interaction. 
Given two layers $L_1$ and $L_2$ and a time lag $\Delta y$ we compute a signal-to-noise ratio $\Phi^{L_1\to L_2}(\Delta y)$ 
as the average for different years $y$ of the fraction of significant links $a_1\to a_2$ in the matrix $B_{a_1\to a_2}^{L_1\to L_2}(y,y+\Delta)$.
We consider the links significantly more co-occurring than random with a 99\% confidence level. 
Therefore for two unrelated layers we expect $\Phi\simeq 1\%$, and
any value above that is the footprint of signal overcoming the noise. 
Repeating this simple exercise for different temporal windows allows shedding light on the following issue: 
how many years does it take on average for the structure of activities in $L_1$ to influence the activities in $L_2$? 

\begin{figure}
\begin{center}
\includegraphics[width=0.948\textwidth]{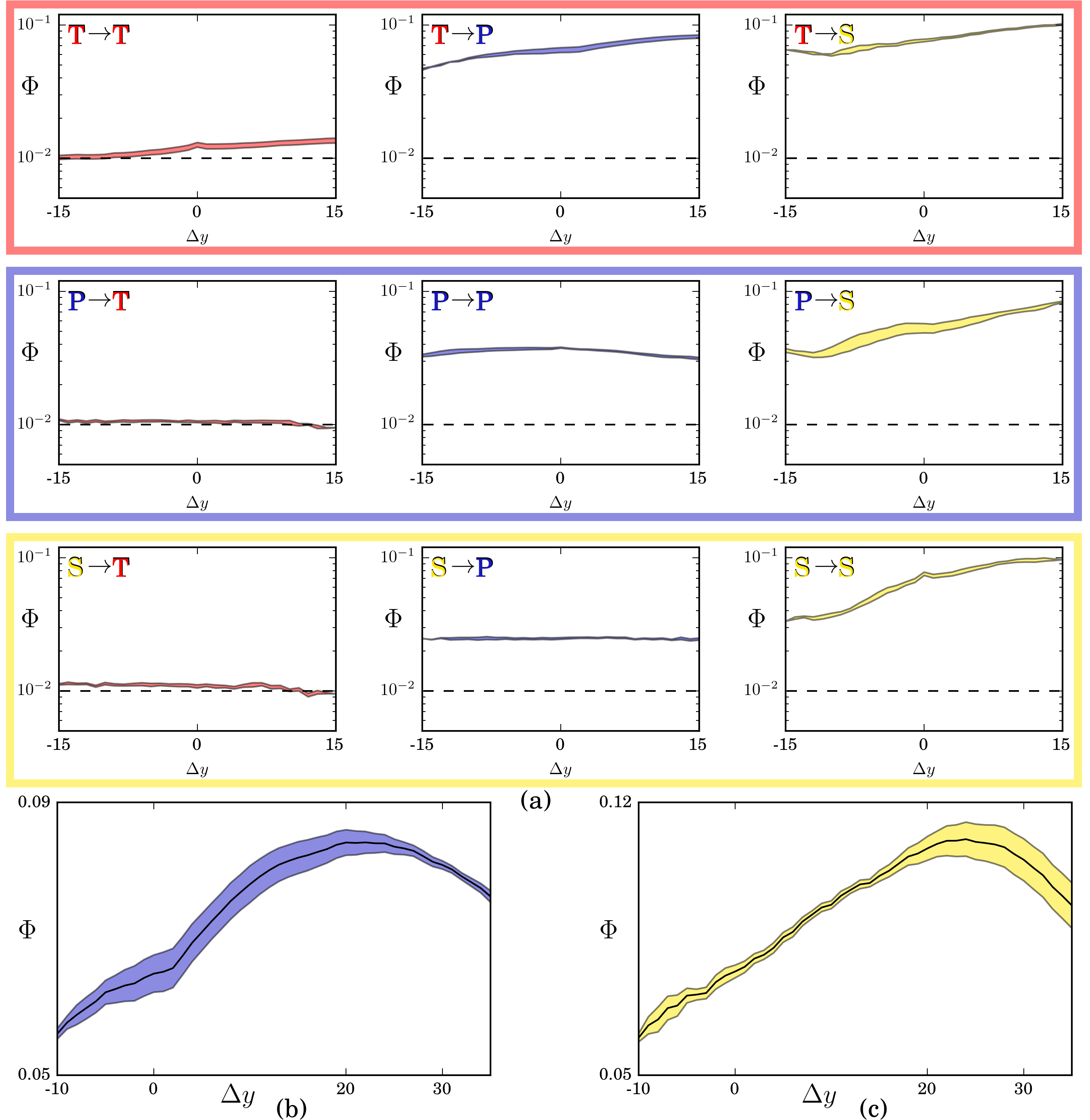}
\caption{(a) Signal $\Phi$ of activities related to \emph{T}echnology, \emph{P}roducts and \emph{S}cience (in this order) for varying time differences $\Delta y$. 
Each plot represents the signal $\Phi$ found when we look at the effect of the activities of the layer represented by the row on the activities of the layer defined by the column.
The time series are build aggregating three years at a time and looking at all the possible pair of years giving the desired $\Delta y$. 
The shaded area denote the one sigma confidence interval.
The analysis is done at a medium level of disaggregation: \emph{T}echnology is split in subclasses ($\sim$600), 
\emph{S}cience in categories ($\sim$300) and \emph{P}roduction at 4 digits level ($\sim$1000). 
Dashed black lines mark the noise level $\Phi=1\%$, as we consider significant links at the 99\% confidence interval.
The same analysis with a longer time frame is reported for \emph{T}echnology-\emph{P}roducts (b) and \emph{T}echnology-{S}cience (c) relations.}
\label{fig:timeDep}
\end{center}
\end{figure}

The results, shown in Figure \ref{fig:timeDep}, are striking. 
First, the signal between different layers is (almost everywhere, see ahead) very high even far in time. 
Beyond the slowly-changing structure of countries activity, this is explained with the strong overlap of countries capabilities on different layers. 
For instance, the ability to patent successfully in a given technological field is a strong predictor 
for the successful export of specific products and the publication of papers in specific scientific fields. 
Second, the technology layer is clearly the best precursor for both science and export, and in both cases the signal reaches its maximum after around 20 years: 
knowing a country's preferred technologies today gives the highest predictive power for its preferred scientific fields and market sectors in about two decades.
It is also the layer more difficult to influence: the capabilities in science and technology in a given year do not give any information on the technological activities in the future.
Third, the scientific layer is both the most capability driven, its future activities being strongly related to the present technological, production and even scientific activities, 
and the less informative about the future activities of the country.

Notice however that we are not giving a direct causal interpretation, saying for instance that there is no impact of science on technology. 
In this specific example the easy explanation is that the chance of someone producing a new patent after reading a scientific paper 
is not easily appropriable by the country, while possible advances of science due to new technologies (like the discovery of superconductors after advancements in cryogenic techniques) 
are more likely to be localized in the same country. Moreover, while a technology can have a deep impact on specific scientific fields, scientific research can potentially lead to 
wide technology spillovers in every fields
(think for instance to scientists at CERN inventing the World Wide Web), which however leave no footprint in our signal.
Nevertheless we can confidently say that knowing the technological portfolio of a country gives more information on the future scientific fields than the opposite.

Overall, the methods and techniques we presented -- in addition to shed light on the dynamics of innovation -- can find 
invaluable use to forecast the scientific fields and market sectors in which countries can have (and will have) 
a competitive advantage based on their current patent portfolios, even in the long-term: 
the technical capabilities a country possesses today will define the scientific and market opportunities for the next generation.

\newpage

\section*{\large SUPPLEMENTARY INFORMATION}

\section*{DATASET}

\subsection*{\emph{S}cience}

We use data on scientific productivity and impact of countries collected from the SciVal platform (\url{www.scival.com}), 
a new API aggregating data from Scopus (\url{www.scopus.com}, owned by Elsevier). The database covers journals, trade publications, book series, conference proceedings, and books. 
Note that while Scopus was shown to have a broader coverage than Web of Science, and to have a more reliable classification system than Google Scholar \cite{Waltman2016}, 
analyses based on either of these databases usually yields very similar results---especially when performed at the country level \cite{Archambault2009,Li2010}. 

Collected data cover years from $1996$ to $2013$ and refer to citations accrued by the corpus of scientific publications, each belonging to a given scientific sector (or sub-sector). 
Data are then aggregated at the country level, so that $W_{c,s}^S(y)$ is the number of bibliometric citations obtained overall by the scientific documents 
produced by country $c$ in scientific sector $s$ during year $y$. Note that these values are computed using a full counting method for internationally co-authored papers~\cite{Waltman2016}, 
which may cause a bias towards small countries with high level of internationalization to the detriment of large standalone countries~\cite{Aksnes2012}. 
Biases can arise also for anglophone countries in Social Science and Humanities, as documents written in other languages and published in national journals---which 
are important especially in such sectors---are not covered in full~\cite{Nederhof2006,Sivertsen2012}. However, these sectors are supposedly less connected to patenting and productive activities, 
and as such do not hinder the results of our analyses.

\subsection*{\emph{T}echnology}

We use patent data contained in PATSTAT (\url{www.epo.org/searching-for-patents/business/patstat}) to measure technological capabilities of countries \cite{EPO2014}. 
PATSTAT collects all the patents (approximately 100 millions) by different Patent Offices (almost 100) all around the World. The time span is extremely broad, going from mid-19th century to today. 
Patent Offices organize knowledge by tagging each patent with one or more codes from the International Patent Classification (IPC).
IPC codes define a hierarchical classification consisting of six levels (\emph{sections, sub-sections, classes, sub-classes, groups, sub-groups}), ranging between 8 and over 70 thousand codes. 
PATSTAT also records the country of origin of the applicant (usually a firm) of each patent.
Finally, PATSTAT defines ``families'' of patents according to primary citations among them \cite{OECD2001}, \ie, all patents with common priorities. 
Indeed, multiple patents could be referred to the same innovation, for example because the same firm applied to different patent offices to extend the protection of their patent to wider markets, 
or because specific patent offices have heterogeneous regulation about the limits of one patents.
In particular we use in the following INPADOC families, or extended family, which collects all the documents that are directly or indirectly linked by one or more priorities.

The basic units of observation are thus the families of patents. Each family is related to one or more countries, through the origin of the applicants of the patents in the family, 
and one or more technological code. Of course the technological codes depend on the aggregation used. 
In constructing the matrices, we assume that each family counts as a unit and thus weights accordingly within the matrix. 
Hence, for each family found in our dataset in a given year, we evenly split its unit of weight among all the technology codes and all the countries it maps to. 
With this caveats in mind, we define $W_{c,t}^T(y)$ as the number of patent families, or the attributed parts of such families, in the field $t$ applied by a firm located in country $c$ on year~$y$.

\subsection*{\emph{P}roducts}

To proxy economic production we use the BACI export data, recorded by the UN COMTRADE (\url{https://comtrade.un.org/}) and processed by CEPII \cite{gaulier2010baci}. 
The original database reports the import-export flows among countries with a data span 20 years, from 1995 to 2014. 
This database includes about 5000 products classified according to the Harmonized System 2007 of the World Customs Organization, which denotes them with a set of 6-digits codes organized in a hierarchical way. 
A given code is divided into three 2-digit parts, each specifying one level of the hierarchy. 
Hence, the first part indicates the broadest categories, such as ``live animals and animal products'' (01xxxx) or ``plastics and articles thereof'' (39xxxx). 
The second two digits specify further distinctions in each category, such as ``live swine'' (0103xx) or ``live bovines'' (0102xx), while the last two digit are even more specific. 
The trade flows are quantified in thousands of current US dollars. After a data sanitation procedure, a country-product matrix is obtained, 
whose generic element $W_{c,p}^P(y)$ represents the monetary value of the overall export of country $c$ for product $p$ during year $y$.

\section*{METHODS}

\subsection*{Revealed comparative advantage} 

Given the raw matrices $\{\mathbf{W}^L(y)\}$ for $L\in \{S,T,P\}$ and for the different years $y$, described in the section above, the first task is to determine 
whether a given country $c$ shows a comparative advantage in activity $a$ (belonging to layer $L$), both with respect to other countries as well as to other activities of the same kind. 
This is achieved through the \emph{revealed comparative advantage} ($RCA$) \cite{RCA1965}, an intensive metric computed as the ratio between the weight of activity $a$ in the activity basket of $c$ 
and the weight of activity $a$ in the total world activity. As a comparative advantage is revealed if $RCA>1$, we binarize the raw matrices to obtain new matrices~$\{\mathbf{M}^L(y)\}$:
\begin{equation}
  M_{c,a}^L(y)=
  \left\{\begin{array}{ll}
        1 \qquad\mbox{ if }\quad\dfrac{W_{c,a}^L(y)}{\sum_{a'}W_{c,a'}^L(y)}\Bigg/\dfrac{\sum_{c'}W_{c',a}^L(y)}{\sum_{c',a'}W_{c',a'}^L(y)}\ge1,\\
        0 \qquad\mbox{ otherwise. } 
        \end{array}
  \right.
\end{equation}

The use of the RCA-like metrics is common in studies of scientific, technological and economic activities 
\cite{Bowen1983,Soete1981,Vollrath1991,Archibugi1992,Patel1994,Kim1995,Schubert1989,Costinot2012,Cimini2016,Guevara2016}. 
The use of the same RCA formulation for the different layer here is mainly motivated by having a coherent way to build them. 
Note that the RCA is also properly normalized for the science layer, in order to get rid of the bias towards old papers which had more time to attract citations than recent ones~\cite{Medo2011}.

\subsection*{The multilayer space} 

Once the binary matrices $\{\mathbf{M}^L(y)\}$ are defined, we build the multi-layer space 
connecting productive, technological and scientific activities inspired by the general ideas presented in \cite{Hidalgo2007,Zaccaria2014}. 
Given a pair of layers $(L_1,L_2)\in\{P,T,S\}$, in order to assess whether countries having a comparative advantage in activity $a_1\in L_1$ in year $y_1$ 
are more likely to have an advantage also in activity $a_2\in L_2$ in year $y_2$, we have to perform an appropriate contraction of $\mathbf{M}^{L_1}(y_1)$ with $\mathbf{M}^{L_2}(y_2)$ 
over the country dimension (\ie, the set $C$ of countries), and take the element $(a_1,a_2)$. 
The detailed prescription to build the \emph{Assist} matrix derives from the so called {\em probabilistic spreading} approach \cite{Zhou2007}.

Let us consider a bit of information located on a generic activity $a_1$ in layer $L_1$. We aim at describing how this information can spread to activities in layer $L_2$. 
As first step, information jumps to countries according to the connection patterns of $\mathbf{M}^{L_1}(y_1)$: 
the transition probability that the bit of information goes from $a_1$ to a given country $c$ is $\rho_{a_1\to c}^{{L_1}\to{C}}(y_1)=M_{c,a_1}^{L_1}(y_1)/u_{a_1}^{L_1}(y_1)$, 
where $u_{a_1}^{L_1}(y_1)=\sum_{c\in C}M_{c,a_1}^{L_1}(y_1)$ is the ubiquity (or degree) of $a_1$ in $L_1$ for year $y_1$. We thus assume equal transition probabilities 
for countries having a comparative advantage in $a_1$, which is motivated by the maximum uncertainty principle since we do not want to introduce biases in the processes. 
As second step, information located on countries jumps to activities in layer $L_2$, now following the connection patters of $\mathbf{M}^{L_2}(y_2)$. 
Again assuming maximum uncertainty, the transition probability that the bit of information goes from $c$ to a given activity $a_2$ in layer $L_2$ 
is $\rho_{c\to a_2}^{{C}\to{L_2}}(y_2)=M_{c,a_2}^{L_2}(y_2)/d_c^{L_2}(y_2)$, where $d_c^{L_2}(y_2)=\sum_a M_{c,a}^{L_2}(y_2)$ is the diversification (or degree) of country $c$ in layer $L_2$ for year $y_2$. 
Putting these two steps together, the probability that the bit of information jumps from activity $a_1\in{L_1}$ to activity $a_2\in{L_2}$ finally reads:
\begin{equation}\label{eq.BB}
\sum_{c\in{C}}\rho_{a_1\to c}^{{L_1}\to{C}}(y_1)\rho_{c\to a_2}^{{C}\to{L_2}}(y_2)=
\frac{1}{u_{a_1}^{L_1}(y_1)}\sum_{c\in{C}}{\frac{M_{c,a_1}^{L_1}(y_1) M_{c,a_2}^{L_2}(y_2)}{d_c^{L_2}(y_2)}}\equiv B_{a_1\to a_2}^{L_1\to L_2}(y_1,y_2).
\end{equation}

The above equation defines a bipartite network between layers $L_1$ and $L_2$, which can be interpreted as the flow of information from activities in $L_1$ in year $y_1$ 
to activities in $L_2$ in year $y_2$ (\ie, after a given time). This interpretation is based on the following considerations.
\begin{itemize}
 \item The bit of information (know-how) associated with a generic activity $a_1$ is transferred to the various activities in $L_2$ through the countries having a comparative advantage in $a_1$.
 \item In order to account for the highly competitive nature of countries development dynamics (be it scientific, technological or economic), 
 we can naturally assume that transferring the know-how from $a_1$ to any activity $a_2$ is not convenient for all countries. 
 In particular, a given country $c$ active in $a_1$ will put an effort in such a transfer which is inversely proportional to the number of countries 
 having $a_1$ in their activity baskets. Thus, the probability that country $c$ exploits the information provided by $a_1$ is given by:
 \begin{equation}
 Pr(c|a_1;y_1)\equiv\rho_{a_1\to c}^{{L_1}\to{C}}(y_1).
 \end{equation}
 \item When a country $c$ transfers its know-how in $a_1$ to activities in ${L_2}$, it is natural to assume that a specific activity $a_2$ will be chosen with a probability 
 inversely proportional to the number of activities in ${L_2}$ in which country $c$ is active. This assumption derives from the finite and fixed amount of resources 
 every country is endowed with for activity transfer. Thus, the conditional probability for the transfer of know-how from activity $a_1$ to activity $a_2$ performed by a given country $c$ reads:
 \begin{equation}
 Pr(a_2;y_2|c;a_1;y_1)=Pr(a_2;y_2|c)\equiv\rho_{c\to a_2}^{{C}\to{L_2}}(y_2). 
 \end{equation}
 The latter equality derives from the assumption that the transfer effort to $a_2$ is independent from the starting activity $a_1$ (as well as from $y_1$). 
 In other words, we model the described random jump as a Markov process: conditional on the present state of the system, its future and past states are independent \cite{Rozanov2012}.
\end{itemize}
Finally, the probability composition formula to assess the transition probability from activity $a_1$ to activity $a_2$ leads directly to eq.~(\ref{eq.BB}):
\begin{equation}
Pr(a_2;y_2|a_1;y_1)=\sum_{c\in{C}} Pr(a_2;y_2|c)Pr(c|a_1;y_1)\equiv B_{a_1\to a_2}^{L_1\to L_2}(y_1,y_2).
\end{equation}

Note that the time direction of the process is given by the time lag $y_2-y_1$. In the case $y_1>y_2$, eq. \ref{eq.BB} remains unchanged 
(still representing the transition probability from $a_1$ to $a_2$), but the interpretation in terms of information flows is the opposite: 
$B_{a_1\to a_2}^{L_1\to L_2}(y_1,y_2)$ becomes the probability that a bit of information reaching activity $a_2\in{L_2}$ originally came from activity $a_1\in{L_1}$.

\subsection*{Hypothesis testing for the multilayer space} 

To assess the statistical significance of elements of the Assist matrices, we resort to a null model for the bipartite matrices $\{\mathbf{M}^L(y)\}$, 
built by randomly reshuffling their elements (\ie, the network connections of layer $L$), but preserving country diversifications and activity ubiquities (\ie, degrees). 
This allows to wipe out the signal coming from the network connectivity patters, beyond the information contained into the degree values. 
Yet in order to formalize the null model analytically (and thus avoid relying on a conditional uniform graph test~\cite{Zweig2011,Neal2014}), degree constraints are imposed 
on average---as for the Canonical ensemble in Statistical Mechanics. We thus end up with a null hypothesis described by the {\em Bipartite Configuration Model} (BiCM)~\cite{Saracco2015}, 
an extension of the standard {\em Configuration Model} \cite{Park2004} to bipartite networks.

Formally, the BiCM null model for a given matrix $\mathbf{M}^L(y)$ is defined as the ensemble $\Omega^{L}(y)$ of bipartite network configurations which are 
maximally random, except for the ensemble average of the degrees that are constrained to the values observed in the empirical network: 
$\avg{\tilde{d}_c^{L}(y)}_{\Omega^{L}(y)}=d_c^{L}(y)$ $\forall c\in C$ and $\avg{\tilde{u}_a^{L}(y)}_{\Omega^{L}(y)}=u_a^{L}(y)$ $\forall a\in L$. 
To ease the notation, in the following we omit the explicit dependence of quantities on the layer $L$ and year $y$, which do not vary throughout the construction of the BiCM. 
Furthermore, we use symbols with the tilde for quantities assessed on null model configurations, and without the tilde for observed values.

Let $\tilde{\mathbf{M}}\in\Omega$ be a network configuration in the ensemble and $P(\tilde{\mathbf{M}})$ be the probability of that graph within the ensemble. 
Following the prescriptions from Statistical Mechanics \cite{Huang1987}, the least biased choice of probability distribution is the one that maximizes the Gibbs entropy
\begin{equation}
 S=-\sum_{\tilde{\mathbf{M}}\in\Omega} P(\tilde{\mathbf{M}})\, \ln P(\tilde{\mathbf{M}}), \label{eq:entropy}
\end{equation}
subject to the normalization condition $\sum_{\tilde{\mathbf{M}}\in\Omega} P(\tilde{\mathbf{M}}) = 1$ plus the constraints:
\begin{equation}
 \avg{\tilde{d}_c}_{\Omega} = \sum_{\tilde{\mathbf{M}}\in\Omega} P(\tilde{\mathbf{M}})\,\tilde{d}_c(\tilde{\mathbf{M}}) = d_c\quad\forall c\in C,\qquad 
 \avg{\tilde{u}_a}_{\Omega} = \sum_{\tilde{\mathbf{M}}\in\Omega} P(\tilde{\mathbf{M}})\,\tilde{u}_a(\tilde{\mathbf{M}}) = u_a\quad\forall a\in L. \label{eq:constraints}
\end{equation}
Introducing the respective Lagrange multipliers $\omega$, $\{\mu_c\}_{c\in{C}}$ and $\{\nu_a\}_{a\in{L}}$ (one for each country and activity of the network), 
the probability distribution that maximizes the entropy satisfies, for all configurations $\tilde{\mathbf{M}}\in\Omega$:
\begin{equation}
\begin{split}
0\:=\;& \frac{\delta}{\delta P(\tilde{\mathbf{M}})}\left[S+\omega\left(1-\sum_{\tilde{\mathbf{M}}\in\Omega} P(\tilde{\mathbf{M}})\right)+\right.  \\
& \left.+\sum_{c\in C}\,\mu_c\left(d_c-\sum_{\tilde{\mathbf{M}}\in\Omega}P(\tilde{\mathbf{M}})\,\tilde{d}_c(\tilde{\mathbf{M}})\right)
+\sum_{a\in L}\,\nu_a\left(u_a-\sum_{\tilde{\mathbf{M}}\in\Omega}P(\tilde{\mathbf{M}})\,\tilde{u}_a(\tilde{\mathbf{M}})\right)\right].
\end{split}
\end{equation}
The solution is:
\begin{equation}
P(\tilde{\mathbf{M}}\,|\,\{\mu_c\},\{\nu_a\})=e^{-H(\tilde{\mathbf{M}}\,|\,\{\mu_c\},\{\nu_a\})}\Big/Z(\{\mu_c\},\{\nu_a\}), \label{eq:p_good}
\end{equation}
where $H(\tilde{\mathbf{M}}\,|\,\{\mu_c\},\{\nu_a\})$ is the graph Hamiltonian and $Z(\{\mu_c\},\{\nu_a\})$ is the partition function 
\begin{equation}
H(\tilde{\mathbf{M}}\,|\,\{\mu_c\},\{\nu_a\})=\sum_{c\in C}\mu_c\,\tilde{d}_c(\tilde{\mathbf{M}})+\sum_{a\in L}\nu_a\,\tilde{u}_a(\tilde{\mathbf{M}}),\label{eq:acca}
\end{equation}
\begin{equation}
Z(\{\mu_c\},\{\nu_a\})=e^{\omega+1}=\sum_{\tilde{\mathbf{M}}\in\Omega} e^{-H(\tilde{\mathbf{M}}\,|\,\{\mu_c\},\{\nu_a\})}. \label{eq:zeta}
\end{equation}
Equations (\ref{eq:p_good}), (\ref{eq:acca}) and (\ref{eq:zeta}) define the BiCM model, namely the distribution over a specified set of network configurations 
that maximizes the entropy subject to the known constraints. As we are considering local constraints (the degrees), 
we can work out on eq. (\ref{eq:p_good}) to obtain \cite{Saracco2015}:
\begin{equation}
P(\tilde{\mathbf{M}}\,|\,\{\mu_c\},\{\nu_a\})=\prod_{c\in C}\prod_{a\in L} \pi_{c,a}^{\tilde{M}_{c,a}}\,(1-\pi_{c,a})^{\tilde{M}_{c,a}}, \label{eq:p_expl}
\end{equation}
where $\pi_{c,a}$ is the ensemble probability for the connection between country $c$ and activity $a$:
\begin{equation}
 \pi_{c,a}=\avg{\tilde{M}_{c,a}}_{\Omega}=\sum_{\tilde{\mathbf{M}}\in\Omega} \tilde{M}_{c,a}\,P(\tilde{\mathbf{M}}\,|\,\{\mu_c\},\{\nu_a\})=\frac{\eta_c\,\theta_a}{1+\eta_c\,\theta_a} \label{conn_prob}
\end{equation}
with $\eta_c=e^{-\mu_c}$ and $\theta_a=e^{-\nu_a}$. Note that in eq. (\ref{eq:p_expl}) the network probability is obtained as the product of connection probabilities over all possible country-activity pairs, 
meaning that in the BiCM context links results as independent random variables. 
The probability distribution in eq. (\ref{eq:p_expl}) yet depends on the values of the Lagrange multipliers, which have to be estimated as:
\begin{equation}
-\frac{\partial}{\partial \mu_c} \ln Z(\{\mu_c\},\{\nu_a\})=\avg{\tilde{d}_c}_{\Omega}\quad\forall c\in C,
\qquad -\frac{\partial}{\partial \nu_a} \ln Z(\{\mu_c\},\{\nu_a\})=\avg{\tilde{u}_a}_{\Omega}\quad\forall a\in L.
\end{equation}
To obtain the numerical value of the ensemble average of the constraints, we maximize the log-likelihood function:
\begin{equation}
\mathcal{L}(\{\mu_c\},\{\nu_a\})=\ln P(\mathbf{M}\,|\,\{\mu_c\},\{\nu_a\})=
\sum_{c\in C}d_c\,\ln\eta_c+\sum_{a\in L}u_a\,\ln\theta_a-\sum_{c\in C}\sum_{a\in L}\ln(1+\eta_c\,\theta_a), \label{eq:likelihood}
\end{equation}
which means solving the system of $|C|+|L|$ equations in $|C|+|L|$ unknowns:
\begin{equation}
  \left\{\begin{array}{ll}
        d_c = \sum_{a\in L} \pi_{c,a} = \sum_{a\in L} \dfrac{\eta_c\,\theta_a}{1+\eta_c\,\theta_a} \qquad\forall c\in C\\
        u_a = \sum_{c\in C} \pi_{c,a} = \sum_{c\in C} \dfrac{\eta_c\,\theta_a}{1+\eta_c\,\theta_a} \qquad\forall a\in L
        \end{array}
  \right.
\end{equation}
Connection probabilities of eq. (\ref{conn_prob}) are now well defined, and can be used to directly sample the ensemble of bipartite configurations or to compute the quantities of interest analytically.

Reintroducing in the notation the explicit dependence on the layer $L$ and year $y$, we finally build the null model $\Omega^{L_1\to L_2}(y_1,y_2)$ for the Assist matrix 
from layer $L_1$ at $y_1$ to layer $L_2$ at $y_2$. This is done by contracting the two BiCMs for the matrices 
$\mathbf{M}^{L_1}(y_1)$ and $\mathbf{M}^{L_2}(y_2)$ along the country dimension~\cite{Gualdi2016,Saracco2016}, as for eq.~(\ref{eq.BB}). We have:
\begin{equation}
\tilde{B}_{a_1\to a_2}^{L_1\to L_2}(y_1,y_2)=\frac{1}{\tilde{u}_{a_1}^{L_1}(y_1)}\sum_{c\in{C}}{\frac{\tilde{M}^{L_1}_{c,a_1}(y_1)\tilde{M}^{L_2}_{c,a_2}(y_2)}{\tilde{d}_c^{L_2}(y_2)}}.
\label{assitz}
\end{equation}

The distributions of $\tilde{B}_{a_1\to a_2}^{L_1\to L_2}(y_1,y_2)$ values describing the null model can be in principle obtained using numerical~\cite{Gualdi2016} 
or approximated analytical~\cite{Saracco2016} techniques. Due to the non-Gaussianity of such distributions, here we resort to a more practical sampling technique. 
We thus use eqn.~(\ref{conn_prob}) and (\ref{assitz}) to generate null Assist matrix networks, 
and populate the ensemble $\Omega^{L_1\to L_2}(y_1,y_2)$ to estimate the full distributions. 
The generic observed element $B_{a_1\to a_2}^{L_1\to L_2}(y_1,y_2)$ is then deemed statistically significant depending on the $p$-value that we can infer from its distribution under the null hypothesis. 
The specific threshold for statistical significance and the size of the generated ensemble vary on the exercises performed (as highlighted in the text). 
It is useful to recall that the two choices, the threshold and the size of the ensemble, are not unrelated: the higher the threshold we want to test, the bigger the sample we require. 
At the very least, if we set a $99\%$ threshold we naturally need at least $100$ realizations of the null model to distinguish empirical points overcoming the threshold, 
whereas, at least $1000$ realizations would be required to test against a $99.9\%$ threshold. While this bare minimum is not enough if we want to test a specific link, 
it could be enough when, as in Figure 1 of the main text, we just want to check the share of significant links out of a large sample. 
This is indeed the case of the last exercise, whose results are reported in Figure 4 of the main text. 
There we fix a pair of layers $(L_1,L_2)$ and a given aggregation level. Then, for a given time lag $\Delta y$, we set a measure of signal-to-noise ratio $\Phi^{L_1\to L_2}(\Delta y)$ 
equal to average percentage, over the different years $y$, of significant connections observed in the matrix $\mathbf{B}^{L_1\to L_2}(y,y+\Delta y)$ using a threshold of $95\%$. 
Therefore, we expect $\Phi^{L_1\to L_2}(\Delta y)\simeq5\%$ when no signal is found.

\subsection*{Scale of analysis}

The scale of the analysis (\ie, the aggregation level of data) is crucial when performing the various exercises. 
Indeed, even if we do not explicit it in the notation, any analysis can be performed at different aggregation levels, in multiple dimensions. 
\emph{First}, trivially, we can perform the analysis at different aggregations along the different activities. At a very broad aggregation we can consider ``Physics'' as one scientific activity, 
while at a finer aggregation we can consider any subfield of Physics as an activity.
\emph{Second}, we can compute co-occurrences at different geographical aggregations: we can look if two activities co-occur in the same countries, or in the same regions. 
\emph{Third}, we can look at different temporal aggregations: we can compute the raw matrices $\mathbf{W}^{T}(y)$ looking at the patents produced in one month, in one year or in five years.
The choice of the scale of observation can be a relevant ingredient to look at specific effects: for example the capabilities required at the country level to perform an activity, 
like ``diffused security and education'', can be widely different from the capabilities required at the local level, like ``infrastructure'' or a specific ``climatic condition''.
In other situations, the choice of the scale of observation can be driven mostly by our specific interest in a more or less granular result. 
This is for example shown in Figure 3 of the main text, where we look at the technological fields required to kick-start the export of ``Desktop computers'' at different technological resolutions.

There are however practical reasons constraining the possible resolutions which can be accessed. Data availability is indeed a determinant issue. 
For example, while patents can be assigned to a region at any different scale, both by looking at the address of the inventors and at the address of the assignee firm, 
exports are recorded by customs and are therefore not easily available at finer geographical resolutions. In general, if we are looking at the same time at scientific papers, patents, and exports, 
the common geographical resolution cannot be finer than the country, and the time interval cannot be shorter than one year. 
The second constraining reason is the statistical power of our tests. The finer the activity or time resolution, the less the signal to noise ratio is required to validate each link. 
This is due to the fact that there will be less activities of a specific kind in a short time, and therefore randomness can play an important role. 
We have therefore a trade-off, a sort of indetermination principle: if we are interested at specific activities, we have to increase the time window by pooling different years. 
This can be done both by summing up the matrices for the years in the time interval, or by stacking the matrices by considering different yearly observations of the same country as different rows, 
as in \cite{Zaccaria2014}. This is what we do in the paper when we say that we pool different years.
On the contrary, if we are interested at a fine time analysis, like the analysis generating Figure 4 of the main document, either we look at very aggregated fields or, 
like we do in that analysis, we ignore the specific fields and we simply look at the \emph{number} of significant fields.

\end{document}